\documentclass[12pt]{article}
\usepackage{epsf}
\title{ {\bf
The annihilation cross section of dark matter which is driven by
scalar unparticle}}
\author{\vspace{1cm}\\
        {\bf E. O. Iltan}
        \thanks{E-mail address:
        eiltan@metu.edu.tr}
\\
Physics Department, Middle East Technical University \\
        Ankara, Turkey\\ \\
        }

\date{}

\begin{document}
\setlength{\baselineskip}{24pt}
\maketitle
\setlength{\baselineskip}{7mm}
\begin{abstract}
We analyze the annihilation cross section of dark matter which
interacts with the standard model sector over the scalar
unparticle propagator. We observe that the annihilation cross
section of dark matter pair is sensitive to the dark matter mass
and the scaling dimension of scalar unparticle. We estimate a
range for the dark matter mass and the scaling dimension of scalar
unparticle by using the current dark matter abundance.
\end{abstract}
\thispagestyle{empty}
\newpage
\setcounter{page}{1}
The visible matter is considerably less than the amount of matter
required in the universe and $23\%$ of present Universe
\cite{JungmanG, JungmanG2, Clowe, KomatsuE, KomatsuE2} is
contributed by the dark matter (DM) that is not detectable by the
radiation emitted. Although the nature of DM is not known at
present, the weakly interacting massive particles (WIMPs)
\cite{JungmanG} are among the promising candidates of DM and they
are expected in the mass range $10$ GeV- a few TeV. WIMPs do not
decay in to standard model (SM) particles since they are stable,
however they disappear by pair annihilation (see for example
\cite{DEramoF, WanLeiGuo}). One needs a theoretical framework
beyond the SM in order to explain the nature of DM and its
stability that can be ensured by an appropriate discrete symmetry
in various models (for details see for example \cite{ChuanRenChen}
and references therein). From the experimental point of view the
indirect detection of the DM candidate is based on the current
relic density which can be explained by thermal freeze-out of
their pair annihilation. By using the current DM abundance by the
WMAP collaboration \cite{KomatsuE2} one gets the appropriate range
for the annihilation cross section and obtains a valuable
information about the nature of DM. In the present work we take an
additional scalar SM singlet DM field $\phi_S$ (see
\cite{SilveiraV}-\cite{HeXG})  and assume that it interacts with
the SM sector over the scalar unparticle propagator. Unparticles
\cite{Georgi1, Georgi2} arise from the interaction of the SM and
the ultraviolet sector with non-trivial infrared fixed point at
high energy level. They are massless and they have non integral
scaling dimension $d_U$. The unparticle sector weakly interacts
with the SM one and the interactions of unparticles with the SM
fields in the low energy level is defined by the effective
lagrangian
\begin{equation}
{\cal{L}}_{eff}= \frac{\eta}{\Lambda_U^{d_U+d_{SM}-n}}\,O_{SM}\,
O_{U} \,, \label{efflag}
\end{equation}
with the unparticle operator $O_U$, the energy scale $\Lambda_U$,
the space-time dimension $n$ and the parameter $\eta$ which
carries information about the energy scale of ultraviolet sector,
the low energy sector and the matching coefficient
\cite{Georgi1,Georgi2,Zwicky}. In order to formulate the DM
annihilation we start with the effective lagrangian which obeys
the $Z_2$ symmetry $\phi_S\rightarrow -\phi_S$
\begin{eqnarray}
{\cal{L}}_S=\frac{1}{2}\,\partial_\mu\,\phi_S\,\,\partial^\mu\,\phi_S-
\frac{\lambda}{4}\,\phi_S^4-\frac{1}{2}\,m_S^2\,\phi_S^2-
\frac{\lambda_0}{\Lambda_U^{d_U-2}}\phi_S^2\, O_{U} \, ,
\label{Vint}
\end{eqnarray}
where $\lambda_0$\footnote{Notice that we consider $\lambda_0$ as
universal coupling (see for example \cite{Cheung1}), i.e., we take
$\eta=\lambda_0$ and $n=4$ in eq.(\ref{efflag}).} is the effective
coupling which leads to tree level DM-DM-scalar unparticle
interaction. Here the DM scalar $\phi_S$ has no vacuum expectation
value and the $Z_2$ symmetry guarantees the stability of $\phi_S$
which appears as pairs and it can not decay into the SM particles.
On the other hand they are expected to annihilate into SM
particles with the annihilating cross section which obeys the
observed DM abundance. The annihilation process is driven by the
exchange particle(s) and, here, we assume that the scalar
unparticle propagator is responsible for this annihilation. The
scalar unparticle propagator is obtained by using the scale
invariance \cite{Georgi2, Cheung1}:
\begin{eqnarray}
\!\!\! \int\,d^4x\,
e^{ipx}\,<0|T\Big(O_U(x)\,O_U(0)\Big)0>=i\frac{A_{d_u}}{2\,\pi}\,
\int_0^{\infty}\,ds\,\frac{s^{d_u-2}}{p^2-s+i\epsilon}=i\,\frac{A_{d_u}}
{2\,sin\,(d_u\pi)}\,(-p^2-i\epsilon)^{d_u-2} \, ,
\end{eqnarray}
where $A_{d_u}=\frac{16\,\pi^{5/2}}{(2\,\pi)^{2\,d_u}}\,
\frac{\Gamma(d_u+\frac{1}{2})} {\Gamma(d_u-1)\,\Gamma(2\,d_u)}$
and the function $\frac{1}{(-p^2-i\epsilon)^{2-d_u}}$ becomes $
\frac{1}{(-p^2-i\epsilon)^{2-d_u}}\rightarrow
\frac{e^{-i\,d_u\,\pi}}{(p^2)^{2-d_u}}$ for $p^2>0$ with a
non-trivial phase which appears as a result of non-integral
scaling dimension.

The total averaging annihilation rate of DM can obtained by the
process $\phi_S\,\phi_S\rightarrow U \rightarrow X_{SM}$,
\begin{eqnarray}
<\sigma\,v_r>&=&\frac{4\,\lambda_0^2}{m_S\,\Lambda_U^{2\,(d_U-2)}}\,
\Bigg(\frac{A_{d_U}}{2\,sin\,d_U\,\pi}\,\Big(\frac{1}{4\,m_S^2}
\Big)^{2-d_U}\Bigg)^2\, \Gamma(\tilde{U}\rightarrow X_{SM})\, ,
\label{sigmavr}
\end{eqnarray}
where $\Gamma(\tilde{U}\rightarrow
X_{SM})=\sum_i\,\Gamma(\tilde{U}\rightarrow X_{i\,SM})$ with
virtual unparticle $\tilde{U}$ having mass $m_U=2\,m_S$ (see
\cite{BirdC, BirdC2}) and $v_r=\frac{2\,p_{CM}}{m_S}$ is the
average relative speed of two DM scalars (see for example
\cite{HeXG}). At this stage we present the functions
$\Gamma(\tilde{U}\rightarrow X_{i\,SM})$ which appear in the
annihilation cross section arising from possible annihilations
that are valid for the DM mass range we choose (see Discussion
section): In this range the annihilations to the
fermion-antifermion pairs\footnote{The annihilations into top-
antitop quark pair and top quark- antineutrino do not exist.},
photon pair, gluon pair and $WW^*$, $ZZ^*$ can exist. For the
fermion-antifermion output we have
\begin{eqnarray}
\Gamma(\tilde{U}\rightarrow
f\,\bar{f})=\sum_f\frac{N_f\,(c_U^{ff})^2} {8\,\pi\,m^2_U}
\,(m_U^2-4\,m_f^2)^{\frac{3}{2}} \, , \label{DWff}
\end{eqnarray}
where $N_f=1\,(3)$ for leptons (quarks) and
$c_U^{ff}=\frac{\lambda_0}{\Lambda_U^{d_U-1}}$. The one for the
photon-photon (gluon-gluon) pair reads
\begin{eqnarray}
\Gamma(\tilde{U}\rightarrow V\,V)=\frac{\beta\,m^3_U} {64\,\pi}
\,|c_U^{VV}|^2 \, , \label{DWVV}
\end{eqnarray}
where $c_U^{VV}=\frac{4\,i\,\lambda_0}{\Lambda_U^{d_U}}$ and
$\beta=1\,(2)$ for $V=\gamma\,(g)$.  Finally for $WW^*$ and $ZZ^*$
output we get\footnote{Notice that, in this expression, we ignore
the mass of neutrinos.}
\begin{eqnarray}
\Gamma(\tilde{U}\rightarrow W(Z)
W(Z)^*)=\sum_{ij}\,\Gamma_{ij}(\tilde{U}\rightarrow W(Z) W(Z)^*)
\, , \label{DW}
\end{eqnarray}
with
\begin{eqnarray}
\Gamma_{ij}(\tilde{U}\rightarrow W(Z)
W(Z)^*)=\frac{(2\,\pi)^4}{2\,m_U}\,\int\,\delta[P-\sum_{i=1}^3
p_i\,]
 \,\prod_{i=1}^3\,\frac{d^3p_i}{(2\,\pi)^3\,2\,E_i}\,N_f\,|M_{ij}^{W
(Z)}|^2 \, , \label{DWij}
\end{eqnarray}
where $p_i\,(p_j,\,p_3)$ is the outgoing charged lepton or down
quark (incoming neutrino or up quark, outgoing W boson) four
momentum for $\Gamma_{ij}(\tilde{U}\rightarrow W W^*)$, the
outgoing lepton or quark (antilepton or antiquark, outgoing Z
boson) four momentum for $\Gamma_{ij}(\tilde{U}\rightarrow Z
Z^*)$. In eq.(\ref{DWij}) $|M_{ij}^{W}|^2$ reads
\begin{eqnarray}
|M_{ij}^{W}|^2\!\!\!\!\!&=&\!\!\!
\frac{16\,g^2\,c^2_{UW}\,|V_{ij}|^2\,\Big(
(p_i+p_j).p_3\Big)^2}{m_W^6\,\Big(m_W^2-(p_i+p_j)^2\Big)^2}
\,\Bigg\{
2\,m_i^2\,m_j^2\,m_W^2\,(m_i^2+m_j^2-2\,m_W^2)+2\,(m_i^2+m_j^2)\,m_W^2\,
(p_i.p_j)^2\nonumber \\&-& 2\,\Big(
(m_i-m_W)\,(m_i+m_W)\,p_i.p_3+m_i^2\,p_j.p_3 \Big)\,\Big(
m_j^2\,p_i.p_3+(m_j-m_W)\,(m_j+m_W)\,p_j.p_3 \Big)\nonumber \\&+&
p_i.p_j\Bigg((m_i^4+6\,m_i^2\,m_j^2+m_j^4)\,m_W^2-(m_i^2+m_j^2)\,
\Big(2\,m_W^4+\Big((p_i+p_j).p_3\Big)^2\Big) -m_W^6\Bigg)
\Bigg\}\, .
\end{eqnarray}
Here $c_{UW}=\frac{\lambda_0}{\Lambda_U^{d_U}}$, $V_{ij}$ is the
CKM matrix element for up-down quark pairs and $V_{ij}=1$ for
neutrino-charged lepton.
Finally $|M_{ij}^{Z}|^2$ is
\begin{eqnarray}
|M_{ij}^{Z}|^2&=& \frac{32\,g^2\,
c_{UZ}\,\Big((p_i+p_j).p_3\Big)^2}{c_W^2\,m_Z^6\,\Big(m_Z^2-(p_i+p_j)^2
\Big)^2}\Bigg\{(c_L^2+c_R^2)\,\Bigg( 2\,m_i^2\,m_j^2\,\Big(
m_Z^2\,(m_i^2+m_j^2-2\,m_Z^2) \nonumber \\  &-&
\Big((p_i+p_j).p_3\Big)^2\Big) + p_i.p_j\,\Big(
m_Z^2\,\Big(m_i^4+m_j^4-m_Z^4-2\,m_j^2\,(m_Z^2-p_i.p_j)\nonumber\\
&+& 2\,m_i^2\,(3\,m_j^2-m_Z^2+p_i.p_j)\Big)-(m_i^2+m_j^2)\,
\Big((p_i+p_j).p_3\Big)^2\Big)+ 2\,m_Z^2\Big(
m_j^2\,p_i.p_3\,(p_i+p_j).p_3 \nonumber
\\ &+& p_j.p_3\,(-m_Z^2\,p_i.p_3+m_i^2\,(p_i+p_j).p_3 \Big) \Bigg)
-2\,c_L\,c_R\,m_i\,m_j\,\Bigg(m_Z^2\,\Big(m_i^4+m_j^4+3\,m_Z^4
\nonumber\\  &-& 2\,m_j^2\,
(m_Z^2-2\,p_i.p_j)-4\,m_Z^2\,p_i.p_j+4\,(p_i.p_j)^2 +
2\,m_i^2\,(m_j^2-m_Z^2+2\,p_i.p_j) \Big)\nonumber \\&-&
(m_i^2+m_j^2-2\,m_Z^2+2\,p_i.p_j)\,\Big((p_i+p_j).p_3\Big)^2
 \Bigg)\,\Bigg\} \,,
\end{eqnarray}
where $c_{UZ}=\frac{\lambda_0}{\Lambda_U^{d_U}}$,
$c_L=\frac{-1}{2}+s_W^2\,(\frac{1}{2})$ for charged lepton
(neutrino),
$c_L=\frac{-1}{2}+\frac{s_W^2}{3}\,(\frac{1}{2}-\frac{2\,s_W^2}{3})$
for down quark (up quark), $c_R=s_W^2\,(0)$ for charged lepton
(neutrino) and $c_R=\frac{s_W^2}{3}\,(-\frac{2\,s_W^2}{3}$) for
down quark (up quark).-

Now we are ready to analyze annihilation cross section $<
\sigma\,v_r>$ and, by using the expression for the relic
abundance,
\begin{eqnarray}
\Omega\,h^2=\frac{x_f\,10^{-11}\,GeV^{-2}}{< \sigma\,v_r>}
\,,\label{omegahsig}
\end{eqnarray}
with $x_f\sim 25$ \cite{HeXG},
\cite{ServantG}-\cite{Gopalakrishna2}, we get the range
$2.21\times10^{-9}\,GeV^{-2} \leq \,\,< \sigma\,v_r> \,\,\,\leq
2.44\times 10^{-9}\,GeV^{-2}$. Here we respect the upper and the
lower bounds of the present relic abundance \cite{KomatsuE2}
\begin{eqnarray}
\Omega\,h^2=0.1109\pm 0.0056 \, . \label{RelDens}
\end{eqnarray}
\newpage
{\Large \textbf{Discussion}}
\\
In the present work we analyze the annihilation cross section of
DM which interacts with the SM sector over the scalar unparticle
propagator. The DM-DM-unparticle coupling $\lambda_0$ plays an
important role in the annihilation process and we study its
numerical value by respecting the estimated upper and lower bounds
of the annihilation cross section of the DM, namely, $2.21\times
10^{-9}\,GeV^{-2} \leq \,\,< \sigma\,v_r> \,\,\,\leq 2.44\times
10^{-9}\,GeV^{-2}$. Furthermore, the scaling dimension of scalar
unparticle, the energy scale $\Lambda_U$ and the DM mass $m_S$ are
among the free parameters of this scenario. For the the scaling
dimension $d_U$ we choose the well known range $1< d_u <2$ (see
\cite{Georgi2, Liao1}). We consider the DM mass $m_S$ in the
interval $10\, GeV \leq m_S\leq 70\, GeV$ and we take the the
energy scale $\Lambda_U=10\, TeV$.

\begin{figure}[htb]
\vskip -3.3truein \centering \epsfxsize=6.8in
\leavevmode\epsffile{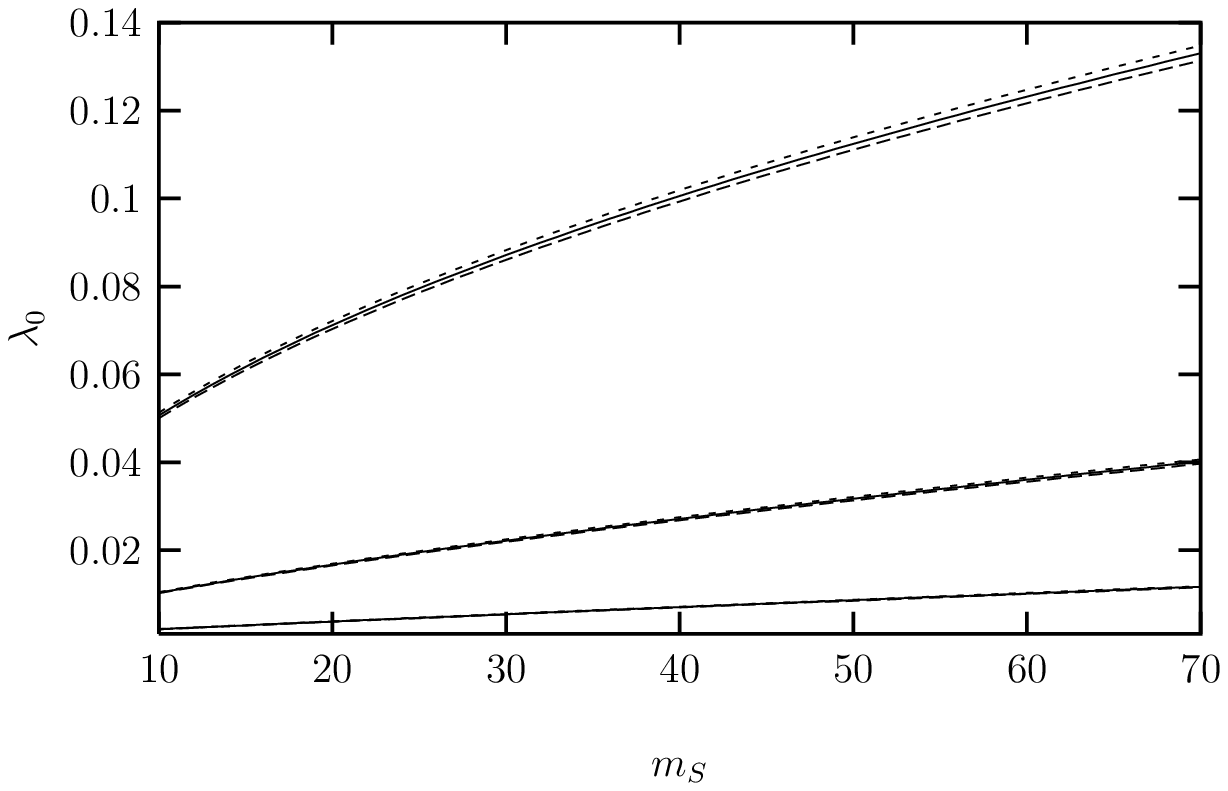} \vskip -3.8truein \caption[]{
$\lambda_0$ as a function of $m_S$. Here the
lower-intermediate-upper solid (long dashed; dashed) line
represents $\lambda_0$ for $d_U=1.1-1.3-1.5$ and $<
\sigma\,v_r>_{AV}$ ($< \sigma\,v_r>_{Max}$; $<
\sigma\,v_r>_{Min}$). } \label{sigmS}
\end{figure}

\begin{figure}[htb]
\vskip -3.3truein \centering \epsfxsize=6.8in
\leavevmode\epsffile{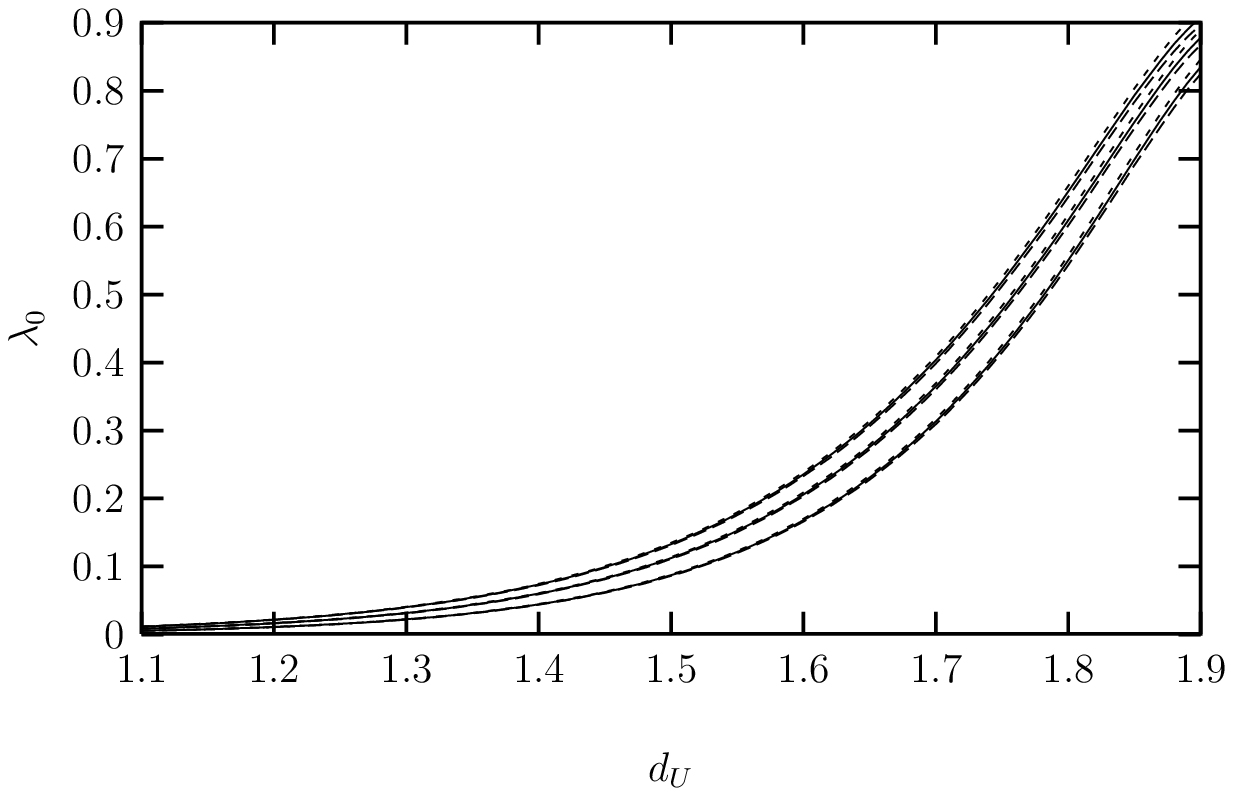} \vskip -3.7truein \caption[]{
$\lambda_0$ as a function of $d_U$. Here the
lower-intermediate-upper solid (long dashed; dashed) line
represents $\lambda_0$ for $m_S=30-50-70\, GeV$ and $<
\sigma\,v_r>_{AV}$ ($< \sigma\,v_r>_{Max}$; $<
\sigma\,v_r>_{Min}$).} \label{sigdu}
\end{figure}

In Fig.\ref{sigmS} we plot the DM mass $m_S$ dependence of the
coupling $\lambda_0$ for the annihilation cross section $<
\sigma\,v_r>$ and different values of $d_U$. Here the
lower-intermediate-upper solid (long dashed; dashed) line
represents $\lambda_0$ for $d_U=1.1-1.3-1.5$ and $<
\sigma\,v_r>_{AV}$ ($< \sigma\,v_r>_{Max}$; $<
\sigma\,v_r>_{Min}$). We observe that the coupling $\lambda_0$ is
sensitive to $m_S$ and this sensitivity increases with the
increasing values of the scaling dimension $d_U$. For small values
of $d_U$ and $m_S$ $\lambda_0$ is more restricted and their
increasing values result in that $\lambda_0$ lies in a broader
range. In order to get the present experimental result of $<
\sigma\,v_r>$, $\lambda_0$ must be at the order of the magnitude
of 0.01 for $1.1< d_u< 1.3$ and it must reach to 0.1 for $d_U=1.5$
for the DM mass values $m_S>40\,GeV$.
For completeness we present the scaling dimension $d_U$ dependence
of the coupling $\lambda_0$ for the annihilation cross section $<
\sigma\,v_r>$ and different values of $m_S$ in Fig.\ref{sigdu}.
Here the lower-intermediate-upper solid (long dashed; dashed) line
represents $\lambda_0$ for $m_S=30-50-70\, GeV$ and $<
\sigma\,v_r>_{AV}$ ($< \sigma\,v_r>_{Max}$; $<
\sigma\,v_r>_{Min}$). This figure also show the strong sensitivity
of the coupling $\lambda_0$ to the scaling dimension $d_U$. The
coupling reaches to the numerical values of the order of $1.0$ for
the upper bounds of  $d_U$, namely for $d_U \sim 1.9$.

\begin{figure}[htb]
\vskip -3.1truein \centering \epsfxsize=6.8in
\leavevmode\epsffile{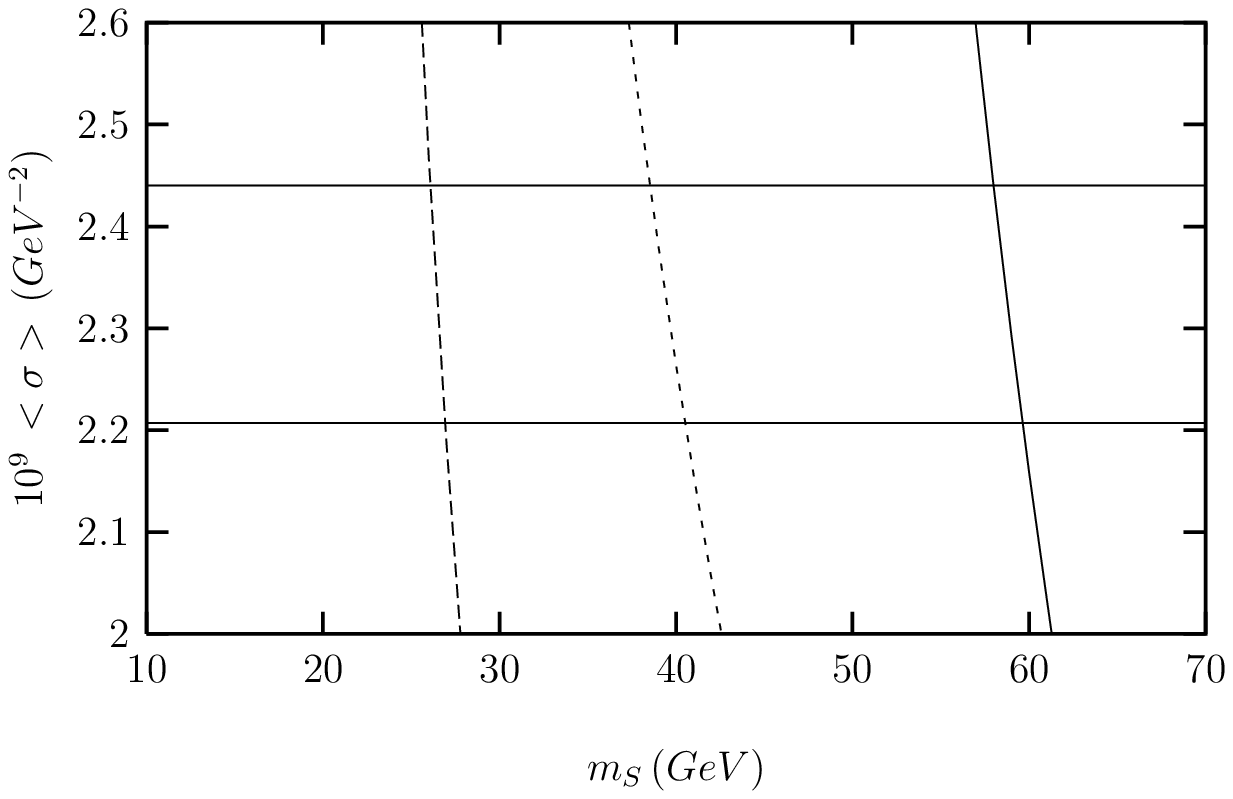} \vskip -3.7truein
\caption[]{The annihilation cross section $< \sigma\,v_r>$ as a
function of $m_S$. Here the solid (long dashed; dashed) line
represents $< \sigma\,v_r>$ for $d_U=1.1$ and $\lambda_0=0.01$
($d_U=1.2$ and $\lambda_0=0.01$; $d_U=1.5$ and $\lambda_0=0.1$). }
\label{crosslam01mS}
\end{figure}

\begin{figure}[htb]
\vskip -3.3truein \centering \epsfxsize=6.8in
\leavevmode\epsffile{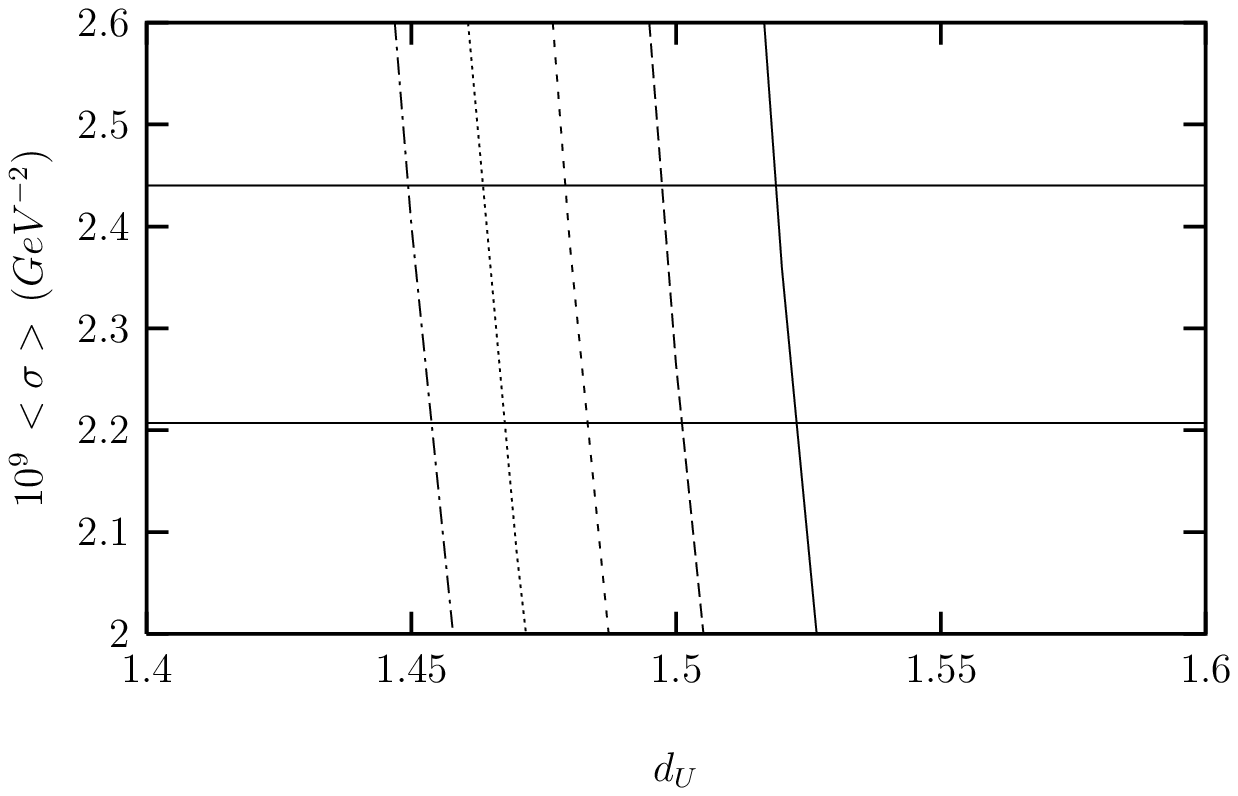} \vskip -3.7truein
\caption[]{The annihilation cross section $< \sigma\,v_r>$ as a
function of $d_U$. Here the solid (long dashed; dashed; dotted;
dot-dashed) line represents $< \sigma\,v_r>$ for $m_S=30\, ( 40;
50; 60; 70)\,GeV$.} \label{crosslam01du}
\end{figure}

Figs.\ref{crosslam01mS} and \ref{crosslam01du} represent $m_S$ and
$d_U$ dependence of the annihilation cross section $<
\sigma\,v_r>$ and, in both figures, the straight solid lines show
the estimated upper and lower bounds.

Fig.\ref{crosslam01mS} is devoted to the annihilation cross
section $< \sigma\,v_r>$ with respect to $m_S$ for different
values of $d_U$ and $\lambda_0$. Here the solid (long dashed;
dashed) line represents $< \sigma\,v_r>$ for $d_U=1.1$ and
$\lambda_0=0.01$ ($d_U=1.2$ and $\lambda_0=0.01$; $d_U=1.5$ and
$\lambda_0=0.1$). We observe that the $< \sigma\,v_r>$ is obtained
in the estimated range for $d_U=1.1$ and $\lambda_0=0.01$ in the
case of $m_S\sim 60\,GeV$. For $d_U=1.2$ and $\lambda_0=0.01$ the
DM mass should be light, namely $m_S\sim 25\,GeV$, to get $<
\sigma\,v_r>$ in the estimated range. For $d_U=1.5$ and
$\lambda_0=0.1$, $< \sigma\,v_r>$ lies in the estimated range for
$m_S\sim 40\,GeV$. We see that, for a fixed coupling $\lambda_0$
(for $\lambda_0=0.01$ see this figure), the increase in the
scaling dimension $d_U$ results in the decrease in the mass $m_S$
so that $< \sigma\,v_r>$ lies in the estimated range.

Fig.\ref{crosslam01du} represents the annihilation cross section
$< \sigma\,v_r>$ with respect to $d_U$ for $\lambda_0=0.1$ and
different values of $m_S$. Here the solid (long dashed; dashed;
dotted; dot-dashed) line represents $< \sigma\,v_r>$ for $m_S=30
\,( 40; 50; 60; 70)\,GeV$. This figure shows that $< \sigma\,v_r>$
lies in the estimated range when $m_S$ respects $30\,GeV < m_S<
70\,GeV$ and $d_U\sim 1.5$.

As a summary, the annihilation cross section $< \sigma\,v_r>$ is
sensitive to the DM-DM-unparticle coupling $\lambda_0$, the DM
mass $m_S$ and the scaling dimension $d_U$. We observe that the
coupling $\lambda_0$ is strongly restricted for the small values
of $d_U$ and $m_S$. The experimental result of $< \sigma\,v_r>$ is
obtained if $\lambda_0$ is at the order of the magnitude of 0.01
(0.1) for $1.1< d_u< 1.3$ ($d_U\sim 1.5$) in the case of
$m_S>40\,GeV$. For $d_U \sim 1.9$, $\lambda_0$ reaches to the
numerical values of the order of $1.0$.

With the forthcoming more accurate experimental measurements one
will provide a considerable information  about the mechanism
driving  the possible annihilation process of DM  and the role of
unparticle physics on this process.
%
%

%
\newpage

\end{document}